\documentclass[aps,twocolumn,epsfig]{revtex4}

\usepackage{graphicx}
\usepackage{psfrag}

\begin{document} 
\title{Instability of capacitance mode 
  in multi-walled nanotubes}
% TU-662
\author{K. Sasaki}
%\email[E-mail: ]{sasaki@tuhep.phys.tohoku.ac.jp}
\affiliation{Department of Physics, Tohoku University, 
Sendai 980-8578, Japan}

\date{\today}

\begin{abstract}
  Conditions for intra and inter layer Coulomb interactions 
  in multi-walled carbon nanotubes are derived from stability 
  of capacitance excitations. It is pointed out, supposing the 
  stability conditions are not satisfied, 
  that the system has unstable modes which correspond to a charge 
  transfer between layers or charge density oscillation in each layer depending
  on zero mode or non-zero modes. 
  It is argued that the stability conditions can be broken when 
  the vacuum polarization processes due to the massive bands are taken into account.
\end{abstract}

\maketitle
%%%%%%%%%%%%%%%%%%%%%%

\section{introduction}

Electrical properties of carbon nanotubes(CNTs~\cite{Iijima,Dekker,Wildoer}) 
have attracted much attention from various points of view. 
They are characterized by the quantum mechanical behavior of the 
$\pi$-electrons which are interacting with each other via the Coulomb interaction.
Several aspects of the Coulomb interaction are already seen
in some experiments of charging~\cite{Tans} 
and of temperature dependent resistivity~\cite{Bockrath}
which are consistent with current understanding of low energy 
excitations in metallic CNTs. The low energy excitations in the
metallic CNTs are theoretically modeled by 
{\it massless} bands (linear dispersion bands)
with the Coulomb interaction~\cite{Kane,EG,YO}.
On the other hand,
the {\it massive} bands (other sub-bands except the massless bands)
modify the Coulomb interaction between massless fermions (electrons in 
the massless bands)
through the vacuum polarization effect~\cite{SFMK}. As for single-wall CNTs
with small diameter, the correction to the long wave length modes of 
the Coulomb interaction is negligibly small.
However in case of multi-walled CNTs, because of their large diameter, 
the gap energy of the massive bands is small; besides, 
the number of the massive bands is large so that 
the correction would be large also. Because of that the low 
energy spectrum of the multi-walled CNTs is thought to be
under the influence of the vacuum polarization due to the massive bands.

The multi-walled carbon nanotubes (MWCNTs) consists of several separate
single-wall CNTs and may have an unique electrical property
which is not shown in each individual single-wall CNT. In this paper, 
we investigate a novel electrical property of a MWCNT which has two layers.
Even in this simple case, there is actually a different 
aspect as compared with the single-wall CNT. That is the
Coulomb interaction between the layers: interlayer Coulomb interaction. 
The Coulomb interactions in this system consist of 
three potentials; intra-layer Coulomb interactions in each layer and
the interlayer interaction. Although it is difficult to obtain information on
the Coulomb potentials as compared with the one in the single-wall CNTs,
it can be thought that the intra-layer Coulomb potential for an inner layer
is the same potential in the single-wall CNTs, 
however the intra-layer one for an outer layer
is not simple because of the screening due to the presence 
of the electrons in the inner layer.
The interlayer unscreened Coulomb potential 
between the layers in the MWCNTs is calculated theoretically~\cite{Egger}.

Since each layer of the MWCNTs has it's own helical structure, we can divide the
MWCNTs into three categories depending on the structural property of each layer.
That are (1) metallic-metallic(M-M) , (2) metallic-semiconducting(M-S) and
(3) semiconducting-semiconducting(S-S) MWCNT. 
If there were no interlayer interaction in the MWCNTs, analysis of 
those system is the same as that of the two individual single-wall CNTs 
and the stability of the system is trivial. But there is 
the interlayer Coulomb interaction. This interaction cause a new excitation mode;
charge transfer between the two layers: capacitance mode which
may cause an instability of the system.
For S-S and S-M types, the stability of the system is thought to be rigid. 
However we need to consider M-M type MWCNTs carefully.
The Coulomb potentials decide the energy spectrum of the capacitance mode.
Hence, for some kinds of potential, the mode could be unstable, that is, it's 
excitation spectrum have a negative eigenvalue. This cause the instability 
of the vacuum of the system and will change the vacuum into 
an another new vacuum state that we do not know within the model used in this
paper. It is the purpose of this article to find conditions on
the stability of the system and check if the condition is satisfied including 
the vacuum polarization correction due to the massive bands.

The paper is composed as follows. A model Hamiltonian for the low energy 
excitations in the M-M type MWCNTs is constructed in Sec.~\ref{sec:model}. 
In Sec.~\ref{sec:zero}, we study the zero mode spectrum of the system
and establish a condition for the stability of the zero mode.
Then we investigate the non zero modes in Sec.~\ref{sec:nozero}
and check if the system is stable in realistic Coulomb interactions.
Summary and discussion are given in Sec.~\ref{sec:conclusion}.

\section{low energy effective model}\label{sec:model}

Because the kinetic terms and the Coulomb interactions 
consist of densities of each layer,
we begin by specifying the densities in both layers.
Let us define the density for the outer layer as $J_{out}$ and
for the inner one as $J_{in}$. Due to the linear dispersion relation near
the Fermi points, the densities are written by the sum of 
left and right current as $J_{out} = J_{L,out} + J_{R,out}$
and $J_{in} = J_{L,in} + J_{R,in}$. Using these definitions of the left and 
right currents, we define a (quantum) field theoretical kinetic Hamiltonian 
respecting the linear dispersion as~\cite{Sasaki}
\begin{eqnarray}
 H_F &=& \Delta \left[ \frac{L}{8} \int_D : J_{L,out}^2 + J_{R,out}^2 : dx 
             -\frac{1}{12} \right] \nonumber \\
 &+& \Delta \left[ \frac{L}{8} \int_D : J_{L,in}^2 + J_{R,in}^2 : dx 
             -\frac{1}{12} \right],
\end{eqnarray}
where $\Delta = 2\pi \hbar v_F/L$ is the single-particle level spacing and is defined 
under the periodic boundary condition along a tubule axis.
$L$ is the length of the tube and $v_F$ is the Fermi velocity.
The integral region is labeled as $D \in [0:L]$.

The densities of each layer are expanded by the current operators as
\begin{eqnarray}
 &&
  J_{out}(x) = \frac{Q_{out}}{L} + \sum_{n\in Z} \left[ (j_L^n)^\dagger + j_R^n  \right] \frac{1}{L} e^{+i2\pi n x/L}, \\
 &&
  J_{in}(x) = \frac{Q_{in}}{L} + \sum_{n\in Z} \left[ (J_L^n)^\dagger + J_R^n  \right] \frac{1}{L} e^{+i2\pi n x/L}.
\end{eqnarray}
The zero modes $Q_{out}$ and $Q_{in}$ denote the total charges in each 
layer respectively. The current operators satisfy the following bosonic 
commutation relations:
\begin{eqnarray}
  &&
  [ j_L^n , (j_L^m)^\dagger ] = 4 n \delta_{nm}, \\
  &&
  [ J_L^n , (J_L^m)^\dagger ] = 4 n \delta_{nm},
\end{eqnarray}
and all of the other commutation relations vanish.
Making use of the operator expression for the charge densities, we
rewrite the kinetic Hamiltonian as the sum of zero mode and non-zero modes,
\begin{eqnarray}
 H_F = H_F^0 + \sum_{n>0} H_F^n,
\end{eqnarray}
where
\begin{eqnarray}
  H_F^0 &=& \frac{\Delta}{16} \left( Q_{out}^2 + Q_{5,out}^2 \right)
  + \frac{\Delta}{16} \left( Q_{in}^2 + Q_{5,in}^2 \right), \\
  H_F^n &=& \frac{\Delta}{4} 
  \left[ (j_L^n)^\dagger j_L^n + (j_R^n)^\dagger j_R^n \right] \nonumber \\ 
  &+& 
  \frac{\Delta}{4} \left[ (J_L^n)^\dagger J_L^n + (J_R^n)^\dagger J_R^n \right].
\end{eqnarray}
The operators $Q_{5,out}$ and $Q_{5,in}$ indicate the zero mode of the current 
in each layer, that is, they are equal to a integral of currents 
$J_{L,out}-J_{R,out}$ and $J_{L,in}-J_{R,in}$ over the nanotube length.

The vacuum of this sector is defined as a direct product of the Dirac seas 
of both layers ($|vac\rangle = |vac,in \rangle \otimes |vac,out \rangle$) 
which are labeled as $|vac,in \rangle$ and $|vac,out \rangle$ that satisfy
the following conditions,
\begin{eqnarray*}
  &&
  J_L^n |vac,in \rangle = J_R^n |vac,in \rangle = 0,  \\
  &&
  j_L^n |vac,out \rangle = j_R^n |vac,out \rangle = 0, 
\end{eqnarray*}
for positive $n$ value.
The vacuum state change into a new vacuum due to 
the intra and inter layer Coulomb interactions.
They are given by
\begin{eqnarray}
 H_C &=& 
  \frac{1}{2} \int \!\!\! \int_D J_{out}(x) V^{out}(x-x') J_{out}(x')dxdx' \nonumber \\
  &+& \frac{1}{2} \int \!\!\! \int_D J_{in}(x) V^{in}(x-x') J_{in}(x')dxdx' \nonumber \\
 &+& \int \!\!\! \int_D J_{in}(x) V(x-x') J_{out}(x')dxdx'.
\end{eqnarray}
The Coulomb potentials for each layer are labeled as $V^{out}$ and $V^{in}$.
Between the layers, there is the Coulomb interaction denoted $V$.
Explicit form of these potentials would be 
difficult to observe experimentally. However we can drive some information 
of them from the stability of the systems, that we will discuss shortly.
The potentials are rewritten by the Fourier series;
\begin{equation}
 V^\alpha(x) = \sum_{n\in Z} \beta^\alpha_n e^{-i 2\pi n x/L},
\end{equation}
where the superscript $\alpha$ is a layer index and 
take an element of the set $\{ in , out ,\  \}$.
The Coulomb interactions are also decomposed into zero mode 
and non-zero modes as
\begin{equation}
 H_C = H^0_C + \sum_{n > 0} H_C^n,
\end{equation}
where the non-zero modes are given by
\begin{eqnarray}
 H^n_C &=& \beta_n^{out} \left[ (j_L^n)^\dagger + j_R^n  \right] 
  \left[ j_L^n + (j_R^n)^\dagger  \right] \nonumber \\
 &+& \beta_n^{in} \left[ (J_L^n)^\dagger + J_R^n  \right] 
  \left[ J_L^n + (J_R^n)^\dagger  \right] \nonumber \\
 &+& \beta_n \left[ (J_L^n)^\dagger + J_R^n  \right] 
  \left[ j_L^n + (j_R^n)^\dagger  \right] \nonumber \\
 &+& \beta_n \left[ J_L^n + (J_R^n)^\dagger  \right]
  \left[ (j_L^n)^\dagger + j_R^n  \right] .
\end{eqnarray}
The Fourier components of the Coulomb interactions $\beta_n^\alpha$
depend on system parameters; diameter and length of CNTs.
Notice that $\beta_n^{in}$ and $\beta_n^{out}$ for $n \ne 0$ are modified by the
vacuum polarization due to the massive bands,
however the zero mode $\beta_0^{in}$ and $\beta_0^{out}$
do not receive any vacuum polarization correction.
Standard Tomonaga-Luttinger theories of nanotubes take into account
effects of non-linear interactions on low energy spectrum with an assumption
$\beta_n^\alpha = \beta_0^\alpha$~\cite{EG,YO}. The Coulomb potentials and 
vacuum polarization depend on the wave number $n$. 
We respect the wave number dependent nature of the Coulomb interactions and  
vacuum polarization effect but do not consider the non-linear interactions 
including a tunneling interaction between layers.

\section{stability condition of zero mode}\label{sec:zero}

The zero mode of the total Hamiltonian is 
\begin{eqnarray}
 H_F^0 + H^0_C &=& \left( \frac{\Delta}{16} + \frac{1}{2} \beta_0^{out} \right) Q_{out} Q_{out} \nonumber \\
 &+& \left( \frac{\Delta}{16} + \frac{1}{2} \beta_0^{in} \right) Q_{in} Q_{in} \nonumber \\
 &+& \beta_0 Q_{in} Q_{out}.
\end{eqnarray}
Suppose that the system is open, that is, 
number of the electron in each layer can be changed freely, 
then the total charges of each layer can take any value. Therefore when 
there is a direction in the space spanned by $(Q_{out},Q_{in})$ 
that the energy of the zero mode become lower,
the system is unstable along that direction.
To avoid such instability, the energy must be elliptic as a function of 
the charges $(Q_{out},Q_{in})$. The condition of elliptic form corresponds to a
negativity of the discriminant and results in an inequality $G_0 < 0$ where
\begin{equation}
  G_0 = \beta_0^2 - \left( \beta_0^{in}+\frac{\Delta}{8} \right)
  \left( \beta_0^{out} + \frac{\Delta}{8} \right).
\end{equation}
It is easy to find out that direction by diagonalizing the zero mode of the total
Hamiltonian, to do so, we define new charges which are given by the following linear
combination of the previous charges in each layer as
\begin{equation}
 \pmatrix{Q_- \cr Q_+} = N
  \pmatrix{ \frac{\beta_0^{out}-\beta_0^{in} - \sqrt{4 \beta_0^2 + (\beta_0^{out} - \beta_0^{in})^2}}{2\beta_0} & 1 \cr \frac{\beta_0^{out}-\beta_0^{in} + \sqrt{4 \beta_0^2 + (\beta_0^{out} - \beta_0^{in})^2}}{2\beta_0} & 1 }
  \pmatrix{Q_{out} \cr Q_{in}},
\end{equation}
where $N$ is a normalization constant. 
Making use of the new charges, we rewrite the zero mode of the Hamiltonian as
\begin{eqnarray}
 && \frac{1}{4}\left( 
  \frac{\Delta}{4} 
  + \beta_0^{out}+\beta_0^{in} 
  - \sqrt{4 \beta_0^2 + (\beta_0^{out} - \beta_0^{in})^2}
\right) Q_-^2 \nonumber \\
 &+&  \frac{1}{4} \left( 
  \frac{\Delta}{4} 
  + \beta_0^{out}+\beta_0^{in} 
  + \sqrt{4 \beta_0^2 + (\beta_0^{out} - \beta_0^{in})^2}
\right) Q_+^2. \nonumber 
\end{eqnarray}
Note that the condition of elliptic is equivalent to the positively of the 
coefficient in front of $Q_-^2$. Physical meaning of the 
new charges $Q_-$ and $Q_+$ is easily recognized in the case of 
$\beta_0^{out} = \beta_0^{in}$. In this case, we have 
\begin{equation}
 \pmatrix{Q_- \cr Q_+} = \frac{1}{\sqrt{2}}\pmatrix{ -1 & 1 \cr 1 & 1 }
  \pmatrix{Q_{out} \cr Q_{in}}.
\end{equation}
Hence, $Q_+$ is a total charge mode and $Q_-$ corresponds to a capacitance charge.
The system is unstable for a moving electron between inner and outer layers
when $G_0 > 0$ and the system is open. 

On the other hand,
when the system is closed, 
the total number of the charge is fixed, then the Hilbert space is
restricted to satisfy the following condition:
\begin{equation}
  \left( Q_{in} + Q_{out} \right) | {\it closed} \rangle = 0.
\end{equation}
The condition of stability of the system turns into a new condition,
\begin{equation}
  \left[ 
    \frac{1}{8} \Delta + 
    \frac{1}{2}\left( \beta_0^{out} + \beta_0^{in} - 2 \beta_0 \right) 
  \right]  > 0.
\end{equation}
It should be noted that the zero mode does not receive the vacuum 
polarization correction. Therefore, the condition for stability must be satisfied
to keep the vacuum stable.
However, the non zero modes are generally receiving the correction and 
depend on dispersion relations of massive bands.

\section{stability condition of non-zero modes}\label{sec:nozero}

We consider non-zero modes of the total Hamiltonian in this section. 
To analyze the energy spectrum of this system, 
first we apply the Bogoliubov transformations 
to the current operators of each layer separately and obtain the following
expression of the Hamiltonian,
\begin{eqnarray}
  H_F^n + H_C^n &=& E_n^{out} 
  \left( (\tilde{j}_L^n)^\dagger \tilde{j}_L^n + (\tilde{j}_R^n)^\dagger \tilde{j}_R^n \right) \nonumber \\
  &+&  E_n^{in} 
  \left( (\tilde{J}_L^n)^\dagger \tilde{J}_L^n + (\tilde{J}_R^n)^\dagger \tilde{J}_R^n \right) \nonumber \\
  &+& a_n \left( \tilde{J}_L^n + (\tilde{J}_R^n)^\dagger \right) \times
  \left( (\tilde{j}_L^n)^\dagger + \tilde{j}_R^n \right) \nonumber \\
  &+& a_n \left( \tilde{j}_L^n + (\tilde{j}_R^n)^\dagger \right) \times
  \left( (\tilde{J}_L^n)^\dagger + \tilde{J}_R^n \right),
\end{eqnarray}
where the transformed current operators are defined as follows,
\begin{eqnarray*}
  &&
  \pmatrix{ \tilde{j}_L^n \cr (\tilde{j}_R^n)^\dagger } = 
  \pmatrix{ \cosh t_n^{out} & \sinh t_n^{out} \cr 
    \sinh t_n^{out} & \cosh t_n^{out} }
  \pmatrix{ j_L^n \cr (j_R^n)^\dagger }, \\
  &&
  \pmatrix{ \tilde{J}_L^n \cr (\tilde{J}_R^n)^\dagger } = 
  \pmatrix{ \cosh t_n^{in} & \sinh t_n^{in} \cr 
    \sinh t_n^{in} & \cosh t_n^{in} }
  \pmatrix{ J_L^n \cr (J_R^n)^\dagger }.
\end{eqnarray*}
From the point of view of the new current operators, the interlayer 
Coulomb interaction is thought to have different coupling as 
compared with the original coupling $\beta_n$ 
that we denote a new interlayer coupling $a_n$:
\begin{equation}
  a_n = \beta_n \left( \cosh t_n^{out} - \sinh t_n^{out} \right)
  \left( \cosh t_n^{in} - \sinh t_n^{in} \right).
\end{equation}
The following Bogoliubov transformation angle and energy spectrum were used to
rewrite the non zero modes of the Hamiltonian,
\begin{eqnarray*}
  && 
  \sinh 2 t_n^\alpha = \frac{\beta_n^\alpha}{E_n^\alpha} , \ \
  \cosh 2 t_n^\alpha = \frac{1}{E_n^\alpha}\left( \frac{\Delta}{4} + \beta_n^\alpha \right),
  \\
  &&
  E_n^\alpha = \frac{\Delta}{4} \sqrt{1+ \frac{8\beta_n^\alpha}{\Delta}}, \ \
  \alpha \in \{ in,out \}.
\end{eqnarray*}

Second we have to diagonalize the interacting terms of both currents whose 
coupling is $a_n$. This can be done by shifting the current operator
as 
\begin{eqnarray}
  &&
  \bar{j}_L^n = \tilde{j}_L^n + \frac{a_n}{E_n^{out}}
  \left( \tilde{J}_L^n + (\tilde{J}_R^n)^\dagger \right), \\
  &&
  \bar{j}_R^n = \tilde{j}_R^n + \frac{a_n}{E_n^{out}}
    \left( (\tilde{J}_L^n)^\dagger + \tilde{J}_R^n \right),
\end{eqnarray}
and the Hamiltonian leads to 
\begin{equation}
  H_F^n + H_C^n =
  E_n^{out} 
  \left( (\bar{j}_L^n)^\dagger \bar{j}_L^n   + (\bar{j}_R^n)^\dagger \bar{j}_R^n  \right)
  + \tilde{H}_n^{in}.
\end{equation}
Here $\tilde{H}^{in}_n$ is defined as
\begin{eqnarray}
&&
 \tilde{H}^{in}_n = E_n^{in} 
  \left( (\tilde{J}_L^n)^\dagger \tilde{J}_L^n + (\tilde{J}_R^n)^\dagger \tilde{J}_R^n \right) \nonumber \\
  &&
  - \frac{2 a_n^2}{E_n^{out}} \left( \tilde{J}_L^n + (\tilde{J}_R^n)^\dagger \right)
  \left( (\tilde{J}_L^n)^\dagger + \tilde{J}_R^n \right). 
\end{eqnarray}
This Hamiltonian implies that an effective attractive modes that 
couple to the Bogoliubov transformed current operators of the inner layer appears. 
The coupling depends on the 
energy spectrum of outer layer. Suppose that the outer layer is the semiconducting 
tube, then the excitation spectrum $E_n^{out}$ is order of the gap energy 
which is very large as compared with the Coulomb energy. Therefore the effective
attractive interaction weakens much.

We further diagonalize the term $\tilde{H}^{in}_n$ as 
\begin{equation}
  \tilde{H}^{in}_n = \bar{E}_n
  \left( (\bar{J}_L^n)^\dagger \bar{J}_L^n + (\bar{J}_R^n)^\dagger \bar{J}_R^n \right),
\end{equation}
where new current operators are defined as
\begin{equation}
\pmatrix{ \bar{J}_L^n \cr (\bar{J}_R^n)^\dagger } = 
  \pmatrix{ \cosh s_n & \sinh s_n \cr 
    \sinh s_n & \cosh s_n }
  \pmatrix{ \tilde{J}_L^n \cr (\tilde{J}_R^n)^\dagger },
\end{equation}
with the definition of the Bogoliubov transformation angle and energy spectrum
\begin{eqnarray*}
  && 
  \sinh 2 s_n = \frac{-\frac{2a_n^2}{E_n^{out}}}{\bar{E}_n} , \ \
  \cosh 2 s_n = \frac{1}{\bar{E}_n}\left( E_n^{in} -\frac{2a_n^2}{E_n^{out}} \right),
  \\
  &&
  \bar{E}_n = E_n^{in} \sqrt{1 - \frac{4a_n^2}{E_n^{out}{E_n^{in}}}}.
\end{eqnarray*}

From the energy eigenvalue of the lowest excitation mode,
we read a condition of stability as
\begin{equation}
   \bar{E}_n > 0,
\end{equation}
which results in the negativity of the discriminant for the n-th Fourier 
component of the Coulomb interactions:
\begin{equation}
    G_n = \beta_n^2 - \left( \beta_n^{in}+\frac{\Delta}{8} \right)
  \left( \beta_n^{out} + \frac{\Delta}{8} \right) < 0.
\end{equation}
This condition of stability is similar to that of the condition of 
stability of the zero mode for open systems. The physical entity of the lowest 
excitation mode is capacitance mode between the layers.
It is easy to prove this fact in the case of $\beta_n^{in} = \beta_n^{out}$.
In this case, total density ($\rho = J_{in} + J_{out}$) and capacitance density
($\sigma = J_{in} - J_{out}$) decouple, so that one can calculate the energy 
spectrum of the capacitance modes without any difficulty. 
The energy spectrum is given by $4nE_n^\sigma$ where 
$E_n^\sigma = \frac{\Delta}{4} \sqrt{1+\frac{8}{\Delta}(\beta_n^{in}-\beta_n)}$
and stability condition of the spectrum reduces to $G_n < 0$.

\begin{figure}[htbp]
  \begin{center}
    \psfrag{a}{$[n]$}
    \psfrag{c}{\hspace{-0.0cm} $\frac{\Delta}{8}$}
    \psfrag{d}{$d=3[{\rm nm}]$}
    \psfrag{e}{$d=5[{\rm nm}]$}
    \psfrag{f}{$d=10[{\rm nm}]$}
    \psfrag{b}{\hspace{-0.8cm} Fourier components $ \beta_n [{\rm meV}]$}
    \includegraphics[scale=0.6]{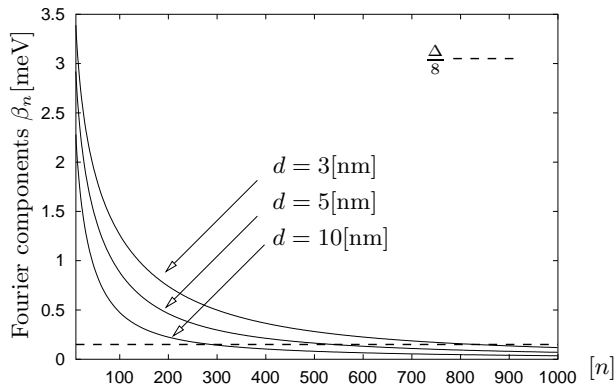}
  \end{center}
  \caption{Fourier components of the interlayer Coulomb interaction
    for several diameters. Horizontal axis denote wave number of the Coulomb
    potential where wave vector is defined by $k_n = 2\pi n/L$. 
    $L$ is the length of the zigzag CNT 
    that we take $L = 3[{\rm \mu m}]$ in these plots.}
  \label{fig:beta}
\end{figure}

We would like to consider several cases and check if the stability 
condition is satisfied.
First, suppose that all the Coulomb potentials
are equivalent, then $\beta_n^{out} = \beta_n^{in} = \beta_n$~\cite{Egger} gives
$G_n < 0$ which means the stability of the system. 
Second, assume the following conditions,
\begin{equation}
 \beta_n = \frac{\beta_n^{out} + \beta_n^{in}}{2},\ \ 
 \beta_n^{out} = \beta_n^{in},
\end{equation}
then 
\begin{equation}
  G_n = -\frac{\Delta}{8} \times 
  \left(  \frac{\Delta}{8} + 2 \beta_n^{in} \right) < 0,
\end{equation}
which also result in the stability of the system.
Finally, we analyze more realistic case.
The diameter of the multi-walled nanotube is thick so that the 
vacuum polarization due to the massive fermions loop can be
no longer negligible and each Fourier component of the intra layer Coulomb potentials
is modified. Therefore the condition of instability is given by
\begin{equation}
  \bar{G}_n = \beta_n^2 - \left( \bar{\beta}_n^{in}+\frac{\Delta}{8} \right)
  \left( \bar{\beta}_n^{out} + \frac{\Delta}{8} \right) > 0,
\end{equation}
where
\begin{equation}
  \bar{\beta}_n^{out} = \frac{\beta_n^{out}}{1-T_n^{out} \beta_n^{out}},\ \ 
  \bar{\beta}_n^{in} = \frac{\beta_n^{in}}{1-T_n^{in} \beta_n^{in}}.
\end{equation}
We denote the static response functions as $T_n^{in}$ and $T_n^{out}$ for each layer.
Here we assume that the interlayer Coulomb interaction does not receive the
vacuum polarization corrections. This assumption is not correct for multi-walled 
CNTs with more than 3 metallic layers. When the vacuum polarization due to the massive
bands is huge, the effective Coulomb couplings $\bar{\beta}_n^{in}$ 
and $\bar{\beta}_n^{out}$ are close to zero. Therefore, the condition of 
instability reduces to the following condition:
\begin{equation}
  \beta_n > \frac{\Delta}{8}.
\end{equation}
We estimate the condition using 
the Fourier mode of the interlayer Coulomb potential which is given by~\cite{EG}
\begin{equation}
  \beta_n = \frac{e^2}{2\pi L} \frac{2}{\pi} \int_0^{\frac{\pi}{2}} 
  K_0 \left( \frac{2n \pi d}{L}\sqrt{\sin^2 x + \left(\frac{a_z}{d}\right)^2 } \right),
\end{equation}
where $K_0$ is the modified Bessel function,
$a_z  = 1.3[\AA]$ is a cutoff length which is introduced to take into account for
the ionization energy of a $\pi$-electron and
$e$ is charge of electron
which gives $e^2/4\pi = 1.44 [{\rm eV \cdot nm}]$.
$d$ should be understood as the mean diameter of the outer and inner layer~\cite{Egger}.
We plot the Fourier components of the Coulomb interaction and the energy 
$\Delta/8$ in Fig.~\ref{fig:beta}. 
Note that there are some instability regions in the spectrum in this case. 

%%%%%%%%%%%%%%%%%%%%%%%%%%%%%
%%%%%%%%%%%%%%%%%%%%%%%%%%%

\begin{figure}[htbp]
  \begin{center}
    \psfrag{a}{$[n]$}
    \psfrag{c}{\hspace{-1.5cm} $\bar{\beta}^{in}_n +\frac{\Delta}{8},d_S$}
    \psfrag{d}{\hspace{-1.5cm} $\bar{\beta}^{in}_n +\frac{\Delta}{8},d_L$}
    \psfrag{e}{$\beta_n, d_S$}
    \psfrag{f}{$\beta_n, d_L$}
    \psfrag{b}{\hspace{-0.7cm}$\beta_n$ and $\bar{\beta}^{in}_n + \frac{\Delta}{8}$ in unit of $[{\rm meV}]$}
    \includegraphics[scale=0.6]{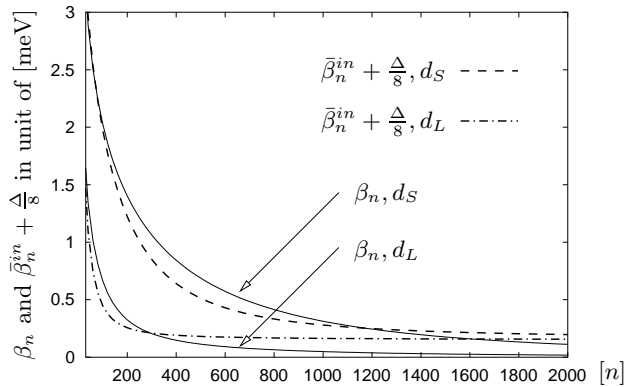}
  \end{center}
  \caption{Fourier components of the interlayer Coulomb interaction
    and of the intra-layer interaction with one-loop vacuum polarization 
    correction plus $\Delta/8$. Horizontal axis denote wave number of the Coulomb
    potential. we take MWCNTs with $L = 3[{\rm \mu m}]$ 
    and $d_S = 1.174,\ d_L=7.045 [{\rm nm}]$.
    Both metallic layers are assumed to be zigzag type CNTs.}
  \label{fig:stability}
\end{figure}

The above estimation in the case of vanishing intra-layer coupling
is a rough estimation. Here we analyze the stability condition 
using a static response function derived at one-loop level~\cite{SFMK}
supposing an equality $\beta_n^{in} = \beta_n^{out} = \beta_n$ and that
$\beta_n^{in}$ and $\beta_n^{out}$ received the same amount of 
the vacuum polarization correction. In this case, the condition of 
instability leads to 
\begin{equation}
  \beta_n > \bar{\beta}_n^{in} + \frac{\Delta}{8}
\end{equation}
We plot $\beta_n$ and $\bar{\beta}_n^{in}+\frac{\Delta}{8}$ for
several MWCNTs with different diameter in Fig.~\ref{fig:stability}.

%%%%%%%%%%%%%%%%%%%%%%%%%%%%%
%%%%%%%%%%%%%%%%%%%%%%%%%%%

\noindent Figure.~\ref{fig:stability} shows that 
there are some unstable spectrum regions in 
momentum space, therefore we conclude that 
the system is thought to be unstable due to the
vacuum polarization of the massive bands. 
Within the unstable region of wave number $n \in U$, the excitation energy 
$\bar{E}_n$ (correctly $4n$ times $\bar{E}_n$) take a negative value. 
Hence the system energy becomes lower if the current operators 
of the capacitance mode have 
a vacuum expectation value.
Although we can not calculate the value of the amplitude,
this instability would cause charge density oscillations in both layers.

\section{summary and discussion}\label{sec:conclusion}

In summary, the stability conditions on the Coulomb
interactions in the multi-walled CNTs consisting of two metallic layers 
have been given within the framework of the simple model 
respecting the quantum nature of the electrons in linear bands 
and the Coulomb interactions.
We have shown that because the massive bands screen the intra-layer 
Coulomb interactions in each layer,
there are finite unstable momentum regions in the system.
Nature of the instability is governed by the capacitance mode 
between layers, therefore it is possible to detect the instability 
experimentally.

The instability conditions are divided into zero and non-zero
modes conditions. 
If the system is unstable, we can not predict a new ground state,
but, can image several candidates of a new vacuum state. 
For example, 
if the zero mode is unstable, charge transfer between the layers is 
expected. After that charge transfer, the electric
state of each layer looks like doped or un-doped single-wall CNTs
(spontaneous capacitance).
As a result of the moving of sufficient amount of charge between layers, 
the system may turn into a stable state
because the low energy excitations are affected by the massive bands.
On the other hand, even in the case of stable zero mode, the non-zero
modes can still be unstable because of the vacuum polarization due to 
the massive bands. In this case, charge density oscillation in both layers
may occur.

We have not considered the tunneling interaction between two layers
in the present paper.
Suppose that such an interaction is present, 
a change of the stability condition is expected. 
We would like to clarify this case in future work.

\begin{acknowledgments}
The author wish to thank A.A. Farajian for fruitful discussion.
\end{acknowledgments}

%%%%%%%%%%%%%%%%%%%%%%%%

\end{document}